\documentclass[iop,apjl]{emulateapj}
\usepackage{apjfonts}

\newcommand{\kms}{km\,s$^{-1}$}

\newcommand{\bz}{$\langle B_{\rm z} \rangle$}
\newcommand{\bs}{$\langle B \rangle$}
\newcommand{\gj}{GJ65}

\newcommand{\fifps}[2]{\centering\resizebox{#1}{!}{\includegraphics{#2}}}

\shorttitle{Magnetic field of GJ65\,A \& B}

\shortauthors{Kochukhov \& Lavail}

\begin{document}

\title{The global and small-scale magnetic fields of fully convective,\\ rapidly spinning M dwarf pair GJ65\,A and B}

\author{Oleg~Kochukhov and Alexis~Lavail}
\affil{Department of Physics and Astronomy, Uppsala University, Box 516, Uppsala 75120, Sweden}

\begin{abstract}
The nearby M dwarf binary GJ65\,AB, also known as BL~Cet and UV~Cet, is a unique benchmark for investigation of dynamo-driven activity of low-mass stars. Magnetic activity of GJ65 was repeatedly assessed by indirect means, such as studies of flares, photometric variability, X-ray and radio emission. Here we present a direct analysis of large-scale and local surface magnetic fields in both components. Interpreting high-resolution circular polarisation spectra (sensitive to a large-scale field geometry) we uncovered a remarkable difference of the global stellar field topologies. Despite nearly identical masses and rotation rates, the secondary exhibits an axisymmetric, dipolar-like global field with an average strength of 1.3~kG while the primary has a much weaker, more complex and non-axisymmetric 0.3~kG field. On the other hand, an analysis of the differential Zeeman intensification (sensitive to the total magnetic flux) shows the two stars having similar magnetic fluxes of 5.2 and 6.7~kG for GJ65\,A and B, respectively, although there is an evidence that the field strength distribution in GJ65\,B is shifted towards a higher field strength compared to GJ65\,A. Based on these complementary magnetic field diagnostic results we suggest that the dissimilar radio and X-ray variability of GJ65\,A and B is linked to their different global magnetic field topologies. However, this difference appears to be restricted to the upper atmospheric layers but does not encompass the bulk of the stars and has no influence on the fundamental stellar properties.
\end{abstract}


\section{Introduction}
\label{intro}

The red dwarf binary system \gj\ consists of two active flare M stars, BL Cet and UV Cet, in a wide orbit ($P_{\rm orb}=26.3$ yr). This binary is a prototype low-mass flare object, repeatedly targeted by photometric, radio, and X-ray observations \citep[e.g.][]{schmitt:2016}. Studies which are able to resolve the \gj\ components indicate that both are M5.5--M6 \citep{henry:1994}, rapidly rotating and spotted \citep{barnes:2016} dwarfs with nearly identical masses and radii \citep{kervella:2016} but a significantly different radio and X-ray behaviour \citep[][and references therein]{audard:2003}.

Here we take advantage of the recent refinement of the fundamental parameters of \gj\ by interferometry and high-contrast imaging \citep{kervella:2016} and an accurate spectroscopic determination of the rotational properties of the components \citep{barnes:2016} to characterise magnetic field strength and topology in both stars. These direct constraints on the surface stellar magnetic field are essential for understanding dynamo action in fully convective stars and interpreting activity patterns of the \gj\ components.

\section{Observational data}
\label{obs}

High spectral resolution circular polarisation observations of \gj\,A and B were obtained with the ESPaDOnS spectropolarimeter attached to the Canada-France-Hawaii Telescope (CFHT). The binary system was observed on the night of Sep 21, 2013, when 4 spectra of \gj\,A and 4 spectra of \gj\,B were recorded, and on the night of Sep 24, 2013, when another 4 spectra of \gj\,B were obtained. According to the orbital parameters derived by \citet{kervella:2016}, the projected separation of the components was 2.2\arcsec\ at the time of ESPaDOnS observations, implying that the stars were well isolated by the 1.6\arcsec\ instrument aperture.

Each spectropolarimetric observation of \gj\,AB consisted of four 480~s sub-exposures obtained with different polarimeter configurations. The resulting Stokes $I$ and $V$ spectra, retrieved from the CFHT Science Archive\footnote{\url{http://www.cadc-ccda.hia-iha.nrc-cnrc.gc.ca/en/cfht/}}, were reduced by the {\sc UPENA} pipeline using the {\sc LIBRE-ESPRIT} software \citep{donati:1997}. These spectra cover the 370--1050~nm wavelength interval at a resolution of about $R=65,000$ and have a peak signal-to-noise ratio (SNR) of 200--300 per pixel. 

The information about individual observations of \gj\,AB are provided in Table~\ref{tbl:obs}, which lists the mid-exposure times, SNRs, and rotation phases calculated with the period of $P=0.2432$~d for \gj\,A and $P=0.2269$~d for \gj\,B \citep{barnes:2016}. 

\begin{table}
\centering
\caption{Journal of spectropolarimetric observations of \gj\,AB.}
\label{tbl:obs}
\begin{tabular}{lllr}
\hline
\hline
HJD & Phase & SNR & \bz\ (G) \\
\hline
\multicolumn{4}{c}{\gj\,A (BL Cet)} \\
2456556.9601 & 0.111 & 314 & $146\pm19$ \\
2456556.9854 & 0.214 & 294 & $167\pm20$ \\
2456557.0628 & 0.533 & 281 & $~41\pm23$ \\
2456557.0883 & 0.638 & 282 & $~37\pm20$ \\
\\[-0.3cm]
\multicolumn{4}{c}{\gj\,B (UV Cet)} \\
2456556.9332 & 0.000 & 260 & $449\pm21$ \\
2456557.0114 & 0.344 & 237 & $555\pm21$ \\
2456557.0368 & 0.456 & 239 & $392\pm22$ \\
2456557.1184 & 0.816 & 246 & $543\pm21$ \\
2456559.9364 & 0.236 & 244 & $660\pm22$ \\
2456559.9943 & 0.491 & 253 & $386\pm19$ \\
2456560.0521 & 0.745 & 244 & $673\pm21$ \\
2456560.1062 & 0.984 & 203 & $412\pm25$ \\
\hline
\end{tabular}
\tablecomments{The rotational phase reported in the second column is computed using the periods from \citet{barnes:2016} and the reference Julian date of the first observation of \gj\,B.}
\end{table}

\section{Magnetic field analysis}
\label{mag}

\subsection{Global magnetic field topology}
\label{bv}

The quality of ESPaDOnS polarisation observations of \gj\ is insufficient for an analysis of polarisation signatures in individual spectral lines. Consequently, we applied the least-squares deconvolution procedure \citep[LSD,][]{donati:1997,kochukhov:2010a} to derive high SNR mean Stokes $V$ profiles based on a line mask containing 1690 atomic absorption lines deeper than 0.2 of the continuum. The LSD atomic line mask was derived from a line list retrieved from the {\sc VALD3} database \citep{ryabchikova:2015} using $T_{\rm eff}=2900$~K, $\log g=5.0$ {\sc MARCS} model atmosphere \citep{gustafsson:2008}. Application of LSD yielded a SNR gain of about 20, allowing to detect circular polarisation signals in all observations of both components.

\begin{figure*}[!th]
\fifps{7.3cm}{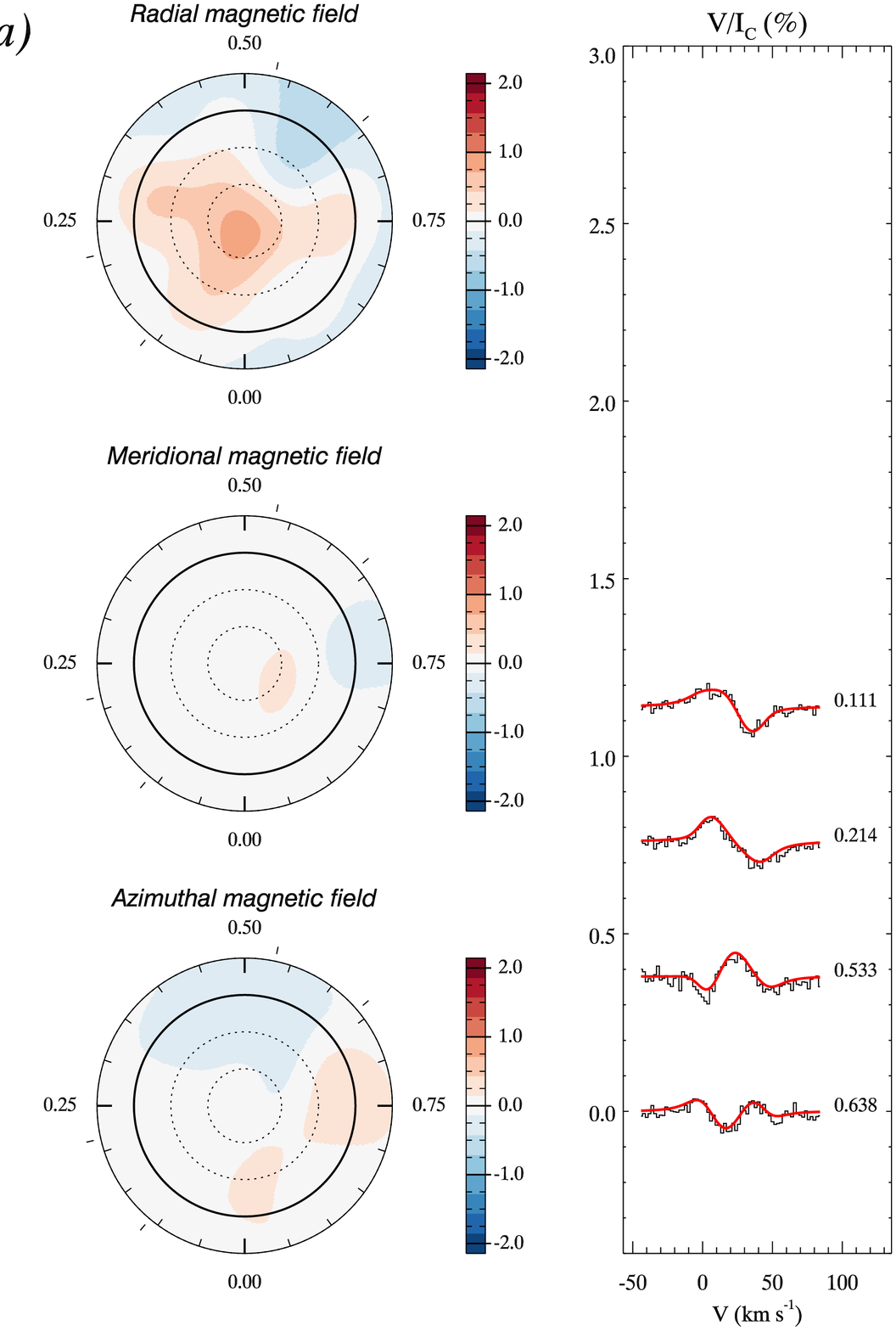}\hspace*{0.3cm}
\fifps{7.3cm}{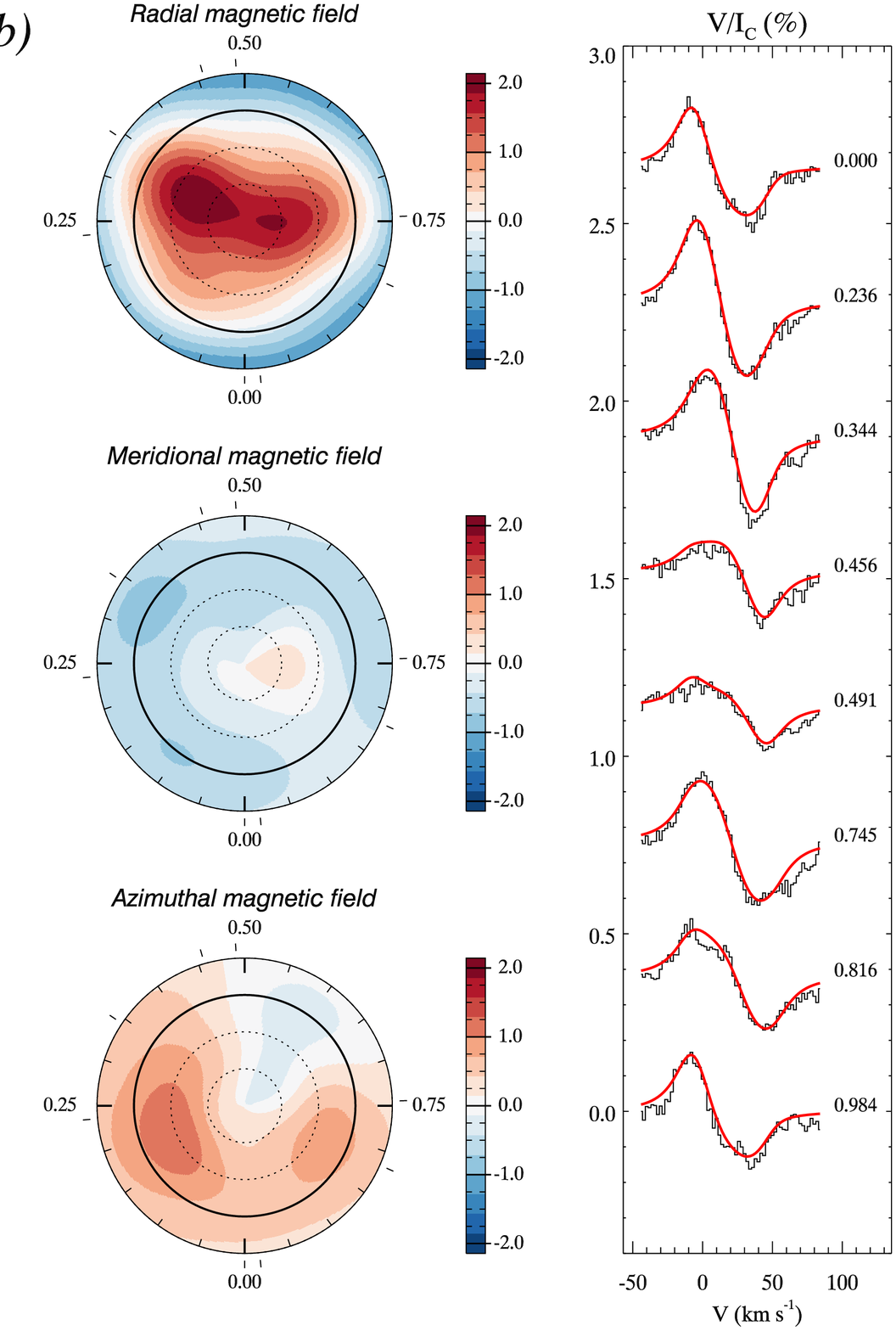}
\caption{
Global magnetic field geometry derived for \gj\,A ({\bf a)} and \gj\,B ({\bf b)} with Zeeman Doppler imaging. For each star the flattened polar projections of the radial, meridional, and azimuthal magnetic field components are presented on the left. The colour bars indicate the field strength in kG. A comparison between the observed (histograms) and model (solid lines) Stokes $V$ profiles are shown on the right. The spectra for different rotational phases are offset vertically.
}
\label{fig:zdi}
\end{figure*}

Examination of the LSD Stokes $V$ profiles of \gj\,A and B reveals a stark difference between the two stars. As demonstrated by Fig.~\ref{fig:zdi}, the primary shows low-amplitude polarisation signatures which change significantly from one observation to the next. On the other hand, the secondary consistently exhibits morphologically simpler and stronger antisymmetric signatures indicative of a positive large-scale magnetic field. The mean longitudinal magnetic field \bz, estimated from the first moment of the Stokes $V$ LSD profiles \citep{wade:2000} and reported in the last column of Table~\ref{tbl:obs}, ranges from $+400$ to $+700$~G for \gj\,B and from 0 to $+150$~G for \gj\,A.

We further analysed the Stokes $V$ LSD profiles of \gj\ with the Zeeman Doppler imaging \citep[ZDI,][]{kochukhov:2016} inversion technique to obtain detailed field topology models. ZDI was previously applied to a sample of active M dwarf stars of different masses by \citet{donati:2008}, \citet{morin:2008,morin:2010}, and \citet{hebrard:2016}. These studies revealed global fields with strengths up to several kG and demonstrated that early-M, partially convective dwarfs typically have weak and complex magnetic field structures while mid-M stars tend to show strong axisymmetric dipolar field topologies, and late-M, fully convective dwarfs exhibit a mixture of both types of field geometries \citep[see review by][]{morin:2012}. 

Our ZDI analysis of \gj\ was carried out with the code described by \citet{kochukhov:2014} and \citet{rosen:2016}. Similar to the latter paper, the local Stokes parameter profiles were approximated with the Unno-Rachkovsky solution of the polarised radiative transfer equations \citep{polarization:2004}. The central wavelength and effective Land\'e factor of the model profiles were adopted according to the average values of the LSD line mask; the line strength was adjusted to match the equivalent width of the observed LSD intensity profile. 

The magnetic field was represented using a spherical harmonic expansion, with the maximum angular degree $\ell_{\rm max}=5$. The magnetic inversions were regularised by penalising a contribution of higher $\ell$ modes \citep[see][]{kochukhov:2014}. This magnetic mapping methodology is very similar to  previous applications of ZDI to active mid- and late-M dwarfs \citep{morin:2008,morin:2010} with the exception that here we do not apply global field filing factors to improve the fit to the observed Stokes $V$ profiles. Furthermore, since observations of \gj\ cover about 10--14\% of the rotation cycles of the components, we accounted for the phase smearing by integrating the model LSD profiles over appropriate phase intervals.

The 8 spectropolarimetric observations of \gj\,B have an adequate phase coverage to obtain a detailed map of the surface magnetic field. On the other hand, the 4 spectra available for \gj\,A cover only about half of the rotational period. Nevertheless, this data set is sufficient to derive an approximate ZDI map that can be meaningfully compared with the inversion results for \gj\,B.

Taking into account results of the Stokes $I$ DI modelling by \citet{barnes:2016}, we adopted projected rotational velocities $v\sin i$\,=\,28.6~\kms\ for \gj\,A and 32.0~\kms\ for \gj\,B, respectively, and an inclination angle of $i=60\degr$ for both stars. This value agrees within error bars with the inclination angles which follow from the above $v\sin i$, the rotation periods determined by \citet{barnes:2016}, and the radii measured by \citet{kervella:2016}.

The global magnetic field maps of \gj\,AB are presented in Fig.~\ref{fig:zdi} along with a comparison of the observed and computed LSD Stokes $V$ profiles. It is evident that the large-scale field of \gj\,B (peak local strength 2.34~kG, mean strength 1.34~kG) is considerably stronger than the field of \gj\,A (peak local strength 0.84~kG, mean strength 0.34~kG). The field of \gj\,B is also predominantly dipolar (92\% of the field energy is contained in $\ell=1$ components) and axisymmetric (89\% of the energy is in $m<\ell/2$ components). In contrast, the field of \gj\,A has a larger contribution of higher-$\ell$ (70\% of the energy is in $\ell=1$ components) and non-axisymmetric (56\% of the energy is in $m<\ell/2$ components) modes. The fields of both stars are largely poloidal (89--93\% of the energy is in poloidal components).

Details of the field topology of \gj\,A are less precisely determined than for the B component due to an incomplete phase coverage obtained for the primary. Nevertheless, the basic conclusion that the primary's field structure must be less axisymmetric and more complex is supported by the complex Stokes $V$ LSD line shapes observed at phases 0.533 and 0.638. Such Stokes $V$ profile morphologies are not observed in any of the 8 observations available for \gj\,B.

\subsection{Total magnetic flux}
\label{bi}

An analysis of Zeeman broadening and intensification of spectral lines enables an estimate of the total magnetic flux, which contains contributions of both the large-scale field and the small-scale magnetic structures unresolved by polarimetry. Previous studies applied this diagnostic method to atomic and molecular lines in the Stokes $I$ spectra of many low-mass stars and brown dwarfs, detecting fields with typical strengths of 2--4~kG \citep{johns-krull:1996,reiners:2007,reiners:2009a,shulyak:2014}. In a recent study by Shulyak et al. (in prep., see also summary in \citealt{kochukhov:2016b}) magnetic fields of up to 6.4~kG were found in several active, rapidly rotating ($P_{\rm rot}$\,=\,0.4--0.8~d) M4--6 dwarfs. The \gj\ components are spinning about a factor of two faster than any of the M dwarf stars with direct magnetic field measurements in the literature.

Our field strength measurements take advantage of a group of \ion{Ti}{1} lines at $\lambda$ 9744--9788~\AA. These lines correspond to the transitions between similar energy levels, hence the relative scale of their oscillator strengths, obtained from {\sc VALD}, is well-known. One of these lines, \ion{Ti}{1} 9743.61~\AA\ has zero effective Land\'e factor, meaning that this spectral feature is entirely insensitive to magnetic field. Then, a magnetic field strength can be determined using a differential spectrum synthesis analysis of magnetically sensitive lines, e.g. \ion{Ti}{1} 9770.30, 9783.31, 9787.69~\AA, relative to \ion{Ti}{1} 9743.61~\AA. This magnetic intensification diagnostic does not rely on interpreting details of the line profile shapes and therefore can be applied to rapidly rotating stars, such as the \gj\ components.

Some of the \ion{Ti}{1} lines mentioned above are significantly affected by the telluric absorption features. We used the {\sc Molecfit} tool \citep{smette:2015} to model the telluric spectrum and remove its contribution from stellar observations. Then, the 4 spectra of \gj\,A and 8 spectra of \gj\,B were averaged, yielding high-quality intensity spectra for each star.

The magnetic field effects on the \ion{Ti}{1} lines were studied with the help of the {\sc Synmast} code \citep{kochukhov:2010a}, which solves the polarised radiative transfer equation at several limb angles for a given line list, model atmosphere, and prescribed magnetic field strength and orientation. In this study we adopted a homogeneous radial magnetic field. This assumption has no bearing on the final analysis results since the disk-integrated Stokes $I$ spectra are not particularly sensitive to the field orientation and, in any case, sample a wide range of field orientations with respect to the observer for any field geometry.

\begin{figure*}[!th]
\fifps{12.5cm}{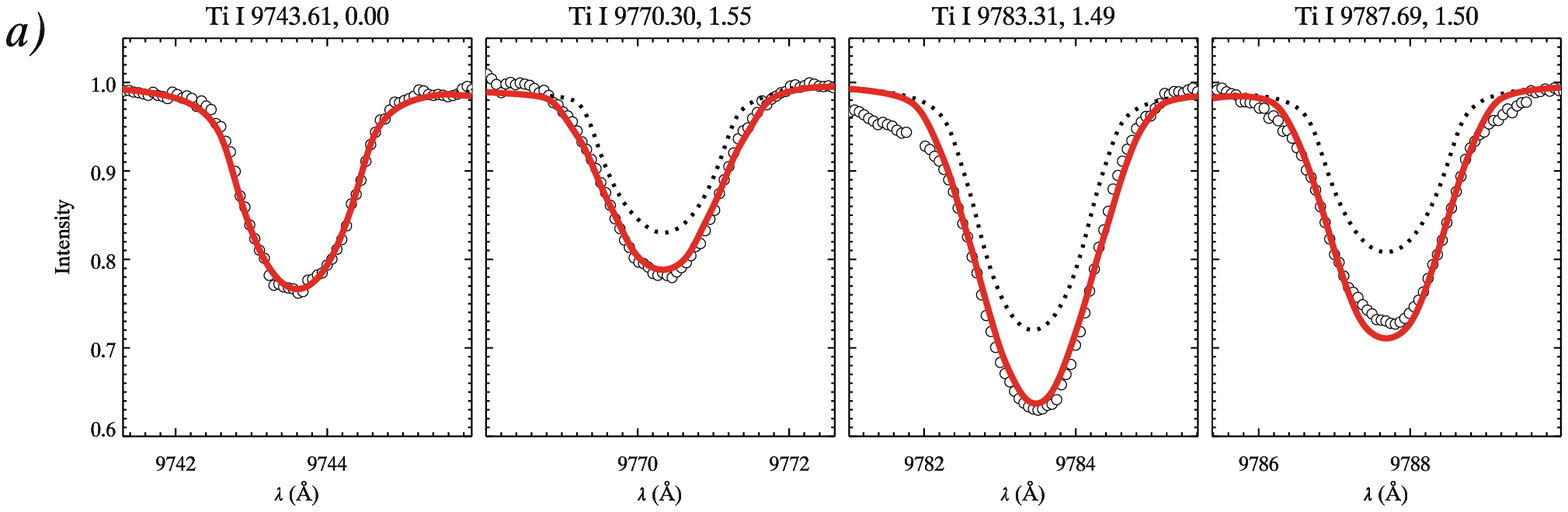}\vspace*{0.3cm}
\fifps{12.5cm}{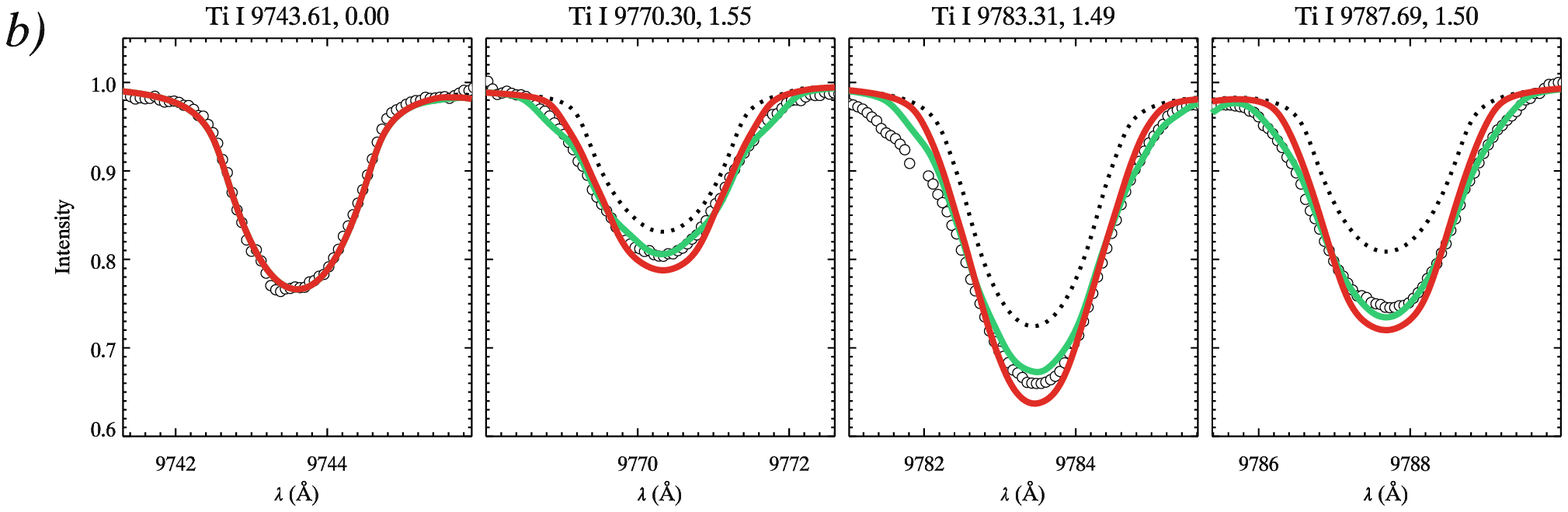}
\caption{
Magnetic intensification and broadening analysis for \gj\,A ({\bf a}) and \gj\,B ({\bf b}) using the four \ion{Ti}{1} spectral lines at $\lambda$ 9743.6--9787.7~\AA. The average observed spectra (open circles) are compared with a non-magnetic synthetic spectrum (dotted line) and with calculations for a 5 kG field covering the entire stellar surface (red line). In addition, the green line in panel {\bf b} illustrates calculations for a 10~kG field covering 60\% of the stellar surface. The central wavelengths of the transitions and their effective Land\'e factors are indicated above each panel.
}
\label{fig:ti}
\end{figure*}

\begin{figure}[!th]
\fifps{8.0cm}{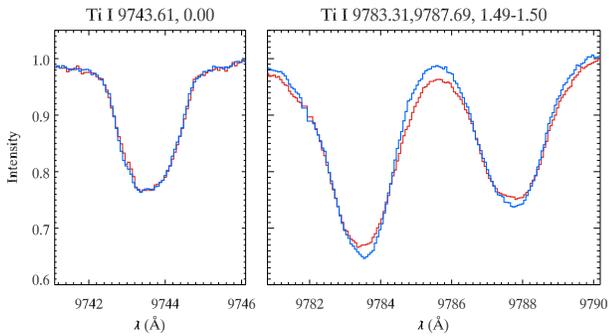}
\caption{
Illustration of the variability of \ion{Ti}{1} lines in \gj\,B. The average over 3 observations in the 0.34--0.49 phase interval (red curve) is compared with the mean of another 3 spectra in the 0.82--1.00 phase interval (blue curve). The left panel shows the magnetic null line \ion{Ti}{1} 9743.61~\AA. The right panel shows two \ion{Ti}{1} lines with Land\'e factors 1.49--1.50.
}
\label{fig:var}
\end{figure}

Based on the radii and masses given by \citet{kervella:2016}, we adopted $\log g$\,=\,5.1 for both components of \gj\ and used the solar-metallicity  {\sc MARCS} model atmospheres with $T_{\rm eff}$\,=\,3000~K for the primary and 2900~K for the secondary. The probable effective temperature uncertainty of $\sim$\,100~K does not introduce significant differential effects for the studied \ion{Ti}{1} lines.

For both \gj\,A and B our spectrum synthesis reproduces the observed width and intensity of \ion{Ti}{1} 9743.61~\AA\ with the solar abundance of titanium and $v\sin i$ given by \citet{barnes:2016}. At the same time, the three other \ion{Ti}{1} lines are clearly too weak without a magnetic field (see Fig.~\ref{fig:ti}). Adjusting the field strength to match the observed intensity of each of these lines separately, we found \bs\,=\,$5.0\pm0.5$~kG for \gj\,A and \bs\,=\,$5.7\pm0.6$~kG for \gj\,B, with the error bars corresponding to the scatter of measurements derived from individual lines.

As illustrated by Fig.~\ref{fig:ti}, theoretical spectrum for a 5~kG field covering the entire stellar surface provides a satisfactory fit to all \ion{Ti}{1} lines in \gj\,A. However, magnetically sensitive lines in \gj\,B exhibit a more triangular shape than predicted by calculations. This anomaly is not observed for the \ion{Ti}{1} 9743.61~\AA\ line in \gj\,B and is not seen for any of the \ion{Ti}{1} lines in \gj\,A. This suggests the presence of a strong-field component in the field distribution in \gj\,B, which is responsible for the extended wings of magnetically sensitive lines. To test this hypothesis, we applied to both stars a two-component model fit, combining a non-magnetic spectrum with calculations for a given field strength covering a fraction $f$ of the stellar surface. This model succeeds in improving the fit to \gj\,B spectra (see Fig.~\ref{fig:ti}), yielding $B$\,=\,$10.5\pm1.2$~kG, $f$\,=\,$0.6\pm0.1$ and \bs\,=\,$B f$\,=\,$6.7\pm0.6$~kG. On the other hand, results of the two-component model fit for \gj\,A ($B$\,=\,$6.0\pm1.5$~kG, $f$\,=\,$0.9\pm0.1$ and \bs\,=\,$B f$\,=\,$5.2\pm0.5$~kG) are compatible with the single-component fit. 

These results were obtained using the mean spectra of \gj\,AB. We have also examined individual observations, finding no appreciable variability of the \ion{Ti}{1} lines in \gj\,A. The secondary component, on the other hand, shows a weak rotational modulation of \ion{Ti}{1} line shapes. This variability, illustrated in Fig.~\ref{fig:var}, appears to be magnetic in nature because it is absent in the magnetic null line \ion{Ti}{1} 9743.61~\AA\ but present in other \ion{Ti}{1} transitions. Fitting the phase resolved spectra of \gj\,B with the two-component model described above reveals a coherent, single-wave variation of $B$ from 8 to 11~kG and \bs\ from 5.7 to 7.0~kG, with the maximum of both parameters occurring at phase 0.5. To our knowledge, this is the first report of a rotational modulation of the magnetic field intensity derived from Stokes $I$ spectra of an M dwarf.

\section{Discussion}
\label{disc}

In this letter we analysed the global magnetic field topology and measured the total magnetic flux for both components of the well-known active M dwarf binary \gj. The components of this system, BL~Cet and UV~Cet, are the fastest spinning fully convective stars for which direct magnetic field measurements are now available. Despite similarity of their fundamental parameters and spin rates, these stars exhibit drastically different global magnetic fields. The secondary, UV~Cet, has a strong, axisymmetric dipolar field. The primary, BL~Cet, has a more complex global field structure, with the magnetic energy more than an order of magnitude weaker than for UV~Cet. This difference in the global field organisation may be responsible for a significantly different radio and X-ray variability of the two stars, with UV~Ceti being both more luminous and variable \citep{audard:2003}. Our results thus suggest that the radio and X-ray behaviour of \gj\,B is linked to an extended magnetosphere anchored in its strong global field, which \gj\,A lacks.

The puzzling finding of very different global fields in nearly identical objects demonstrates that the dynamo in fully convective stars is not a straightforward function of fundamental stellar parameters. The co-existence of different types of large-scale fields in similar late-M dwarfs was previously found by \citet{morin:2010} based on a heterogenous sample of objects with uncertain relative ages and evolutionary histories. Our finding of the discrepant global fields in well-characterised, coeval components of \gj\ strengthens these results and lends support to the hypothesis of convective dynamo bistability \citep{gastine:2013}.

An alternative interpretation of the coexistence of different types of field topologies in M dwarfs with similar fundamental parameters was suggested by \citet{kitchatinov:2014}. These authors argued that M dwarfs have magnetic cycles and observations of qualitatively different field geometries correspond to different phases of these cycles. However, the consistently discrepant radio behaviour of the \gj\ components, which appears to be linked to their different global field configurations, has been observed since the first high angular resolution radio studies of \gj\ \citep{gary:1981}. This implies that any cyclic evolution of the global field in \gj\,AB must occur on a time scale longer than $\sim$\,30 yr.

At the same time, complementary analysis of the total magnetic flux reveals a different perspective on stellar magnetic fields. Similar to previous studies \citep[e.g.][]{reiners:2009}, we find that up to 95\% of the field energy is contained in small-scale field. Moreover, the total mean field strengths derived for \gj\,A and B (5.2 and 6.7~kG respectively) are similar, although there is an evidence of different field strength distributions, with the field in \gj\,B featuring rotationally-modulated, strong-field component absent in \gj\,A. These results, inferred from the analysis of Zeeman intensification, indicate that the discrepant global field topologies diagnosed by polarimetry are limited to superficial stellar layers and immediate circumstellar environment but do not encompass the bulk of the stars. This picture contradicts theoretical bistable dynamo models which predict a different total field strength and different internal field for the two dynamo branches \citep{gastine:2012,gastine:2013}. 

On the other hand, \citet{kervella:2016} reported the same 12--14\% discrepancy between the observed and theoretically computed radii of the \gj\ components. Since the interior magnetic field is the most likely culprit for the inflated radii of active low-mass stars \citep{chabrier:2007,feiden:2012}, \gj\,A and B should have comparable interior fields, which is in line with our finding of the similar total surface magnetic fluxes.

\acknowledgements
The authors acknowledge support by the Knut and Alice Wallenberg Foundation, the Swedish Research Council, and the Swedish National Space Board.

This paper is based on archival spectropolarimetric observations obtained at the CFHT which is operated by the National Research Council of Canada, the Institut National des Sciences de l'Univers (INSU) of the Centre National de la Recherche Scientifique (CNRS) of France, and the University of Hawaii.

%\bibliographystyle{apj}
%\bibliography{astro_papers}

\end{document}